\documentclass{article}
\usepackage{ijcai23}

\usepackage{times}
\usepackage{soul}
\usepackage{url}
\usepackage[hidelinks]{hyperref}
\usepackage[utf8]{inputenc}
\usepackage[small]{caption}
\usepackage{graphicx}
\usepackage{amsmath}
\usepackage{amsthm}
\usepackage{booktabs}
\usepackage{algorithm}
\usepackage{algorithmic}
\usepackage[switch]{lineno}
\usepackage{subfigure}
\usepackage{multirow}

\usepackage{latexsym}

\title{JEPOO: Highly Accurate Joint Estimation of Pitch, Onset and Offset for Music Information Retrieval}

\author{
Haojie Wei$^{1\dagger}$\thanks{Work done when being an intern at Huawei Noah's Ark Lab.}\and
Jun Yuan$^{2}$\thanks{Co-first Authors}\and
Rui Zhang$^{3\dagger}$ \and
Yueguo Chen$^1$\thanks{Corresponding Author}
\and Gang Wang$^2$
\affiliations
$^1$School of Information, Renmin University of China, Beijing, China\\
$^2$Huawei Noah’s Ark Lab, Shenzhen, China\\
$^3$Tsinghua University (\url{www.ruizhang.info})
\emails
\{weihaojie, chenyueguo\}@ruc.edu.cn,
\{yuanjun25, wanggang110\}@huawei.com,
rayteam@yeah.net
}

\begin{document}

\maketitle

\begin{abstract}
    Melody extraction is a core task in music information retrieval, and the estimation of pitch, onset and offset are key sub-tasks in melody extraction. 
    Existing methods have limited accuracy, and work for only one type of data, either single-pitch or multi-pitch. 
    In this paper, we propose a highly accurate method for joint estimation of pitch, onset and offset, named JEPOO. 
    We address the challenges of joint learning optimization and handling both single-pitch and multi-pitch data through novel model design and a new optimization technique named Pareto modulated loss with loss weight regularization.
    This is the first method that can accurately handle both single-pitch and multi-pitch music data, and even a mix of them. 
    A comprehensive experimental study on a wide range of real datasets shows that JEPOO outperforms state-of-the-art methods by up to 10.6\%, 8.3\% and 10.3\% for the prediction of Pitch, Onset and Offset, respectively, and JEPOO is robust for various types of data and instruments.
    The ablation study shows the effectiveness of each component of JEPOO\footnote{This paper has been accepted by IJCAI 2023}. 
    We have made the code of JEPOO available \footnote{\url{https://gitee.com/mindspore/models/tree/master/research/recommend/JEPOO}}.
\end{abstract}

\section{Introduction}\label{sec:intro}
Music information retrieval (MIR) is an essential infrastructure supporting the daily use of the large music platforms.
Melody is critical to music retrieval such as query/search by singing and content based music recommendation. 
Melody is also the core of music understanding as agreed by many existing studies~\cite{hawthorne2017onsets,kim2018crepe,gfeller2020spice,gardner2021mt3,hawthorne2021sequence}. 
However, the vast majority of music data are in their audio form, typically \texttt{.wav} or \texttt{.mp3} files, which do not directly reflect the melody of the music data. 
To obtain the melody of music, we have to perform the so-called \textit{melody extraction}, which converts an audio file into a sequence of notes, consisting of (i) the fundamental frequency f0 (termed \textit{pitch}) of the note, (ii) the start of the note (termed \textit{onset}) and (iii) the end of the note (termed \textit{offset}). 
\begin{figure}[htp]
\centering
\includegraphics[width=0.75\columnwidth]{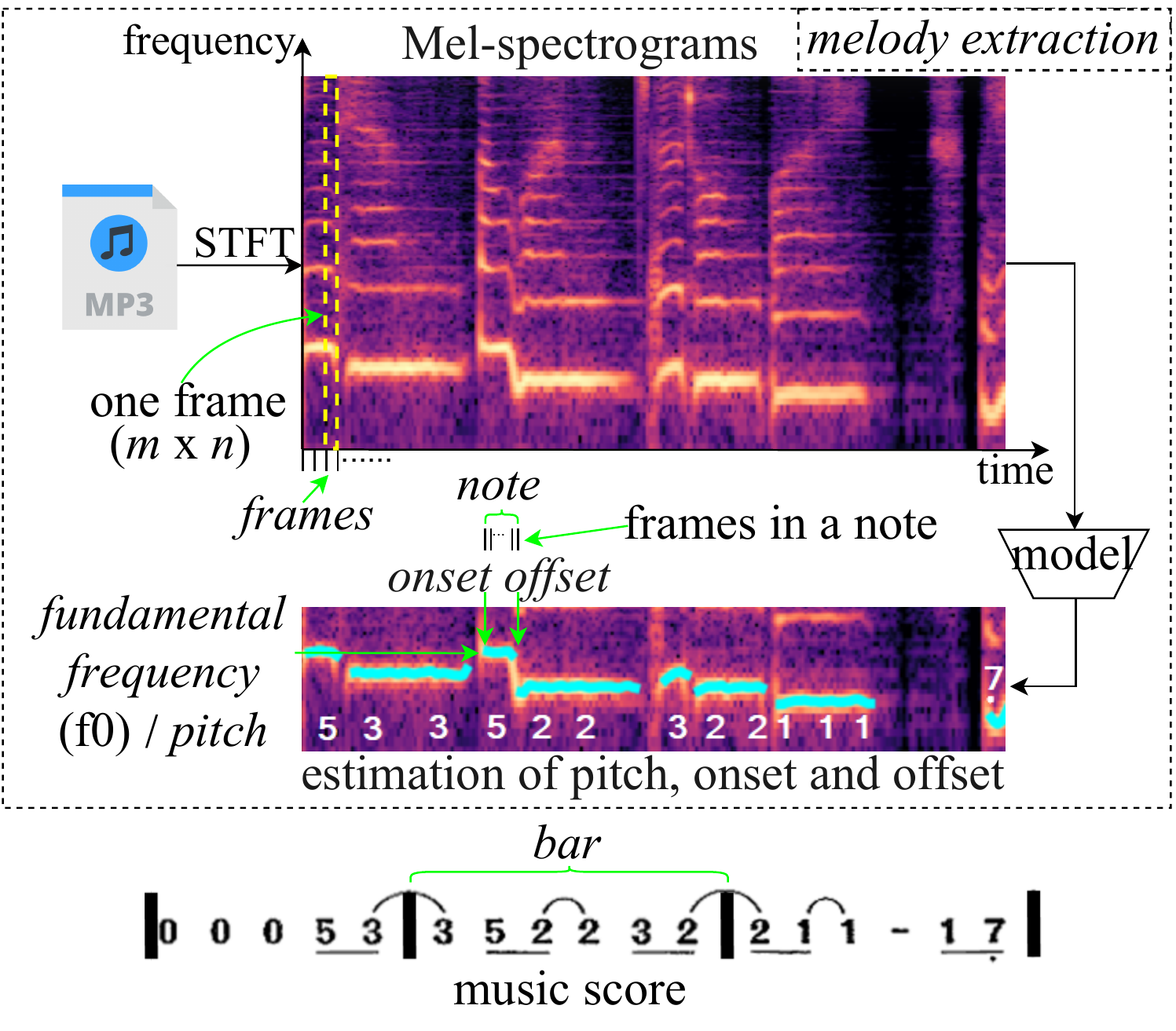}
\caption{Melody extraction. }
\label{fig:melodyextraction}
\end{figure}
Specifically, melody extraction aims to predict the value of f0, the onset, and the offset of each note based on the mel-spectrogram of an audio file as shown in Figure~\ref{fig:melodyextraction}, which is a 2-dimensional matrix denoting the strength of every frequency during every time segment (termed \textit{frame}). 
It is challenging because the boundaries of a note (i.e., its onset and offset) are usually blurry and noisy (c.f. Figure~\ref{fig:melodyextraction}), where the yellow signals tend to have sharp jumps and blurry edges around the beginning and ending of a note.
There has been continued research on melody extraction, but the effectiveness of existing work still needs substantial improvements to be really useful in practice. 
There are two major challenges described below.

\textbf{Challenge 1: Joint learning optimization.} 
One line of studies (e.g., CREPE~\cite{kim2018crepe} and SPICE~\cite{gfeller2020spice}) focus on predicting only pitches but not onsets/offsets. 
The accuracy of these methods is limited and sensitive to noise (c.f. Figure~\ref{fig:case}), because they do not utilize the timing of the notes (onsets/offsets) and do not consider long-term context.
It is reasonable to believe that the prediction of pitch and the prediction of onset/offset should benefit each other. 
Therefore, another line of studies (e.g., OAF~\cite{hawthorne2017onsets} and its follow-ups~\cite{hawthorne2018enabling,kim2019adversarial,kelz2019deep}) perform joint learning of the two tasks \textit{pitches} and \textit{onsets/offsets}. 
However, they perform joint learning by simply summing up the different tasks' losses as the objective function, which do not address the following problems: 
(i) \underline{\textit{Model design}}. 
What is the suitable model structure for the joint prediction of pitch and onset/offset.
(ii) \underline{\textit{Data imbalance}}. 
There is a huge imbalance in the labeled data; Specifically, there are much more negative labels (frames with no pitch) than positive labels (a frame with at least one pitch), and there are even fewer frames with onset/offset labels than those with pitches because a note has only one onset and one offset frame but many frames with pitches (c.f. Figure~\ref{fig:melodyextraction}).
(iii) \ul{\textit{Imbalance between the importance of pitch and onset/offset prediction}}. 
Our ultimate goal is to achieve a balanced accuracy of three tasks, pitch prediction, onset prediction and offset prediction. 
Weighting one task too high may result in high accuracy in one task but low accuracy in the others. 
The optimal weights between the losses of the three tasks may not be the same, which is what existing studies assume when they simply summing up the objective functions of different tasks. 
We need to learn the appropriate weights between the three tasks, which is challenging.

\textbf{Challenge 2: Handling both single-pitch and multi-pitch data.}
Another deficiency of both lines of studies, which seems to be coincidental with the application each line of studies have targeted, is that they have been designed for either single-pitch prediction or multi-pitch prediction. 
This makes them work well for only one case, but not the other case or a mix of both cases. 
Specifically, CREPE~\cite{kim2018crepe}, SPICE~\cite{gfeller2020spice} and their follow-ups have mainly used the single-pitch (\textbf{SP}) dataset such as \texttt{MDB-stem-synth}~\cite{salamon2017analysis}. 
We call them \textit{single-pitch prediction} (\textbf{SPP}) algorithms because they were designed to predict only one pitch at any timestamp.
In comparison, OAF~\cite{hawthorne2017onsets} and its follow-ups~\cite{hawthorne2018enabling,kim2019adversarial,kelz2019deep} focus on multi-pitch (\textbf{MP}) datasets such as \texttt{MAESTRO-V1.0.0}~\cite{hawthorne2018enabling} and \texttt{MAPS}~\cite{emiya2009multipitch}. 
We call them \textit{multi-pitch prediction} (\textbf{MPP}) algorithms because they were designed to predict multiple pitches at any timestamp.

Existing SPP algorithms perform poorly on MP datasets since they are not able to predict multiple pitches at the same timestamp.
Existing MPP algorithms perform poorly on SP datasets because they were trained on MP data and tend to predict multiple pitches in a frame.
Since SPP is a special case of MPP, we may improve MPP algorithms by retraining them on SP data. However, their performance on a mix of SP and MP data is still poor because their decision boundaries are different caused by the positive/negative label imbalance. 
Specifically, SP data has a much lower positive/negative label ratio than that of MP data, because in SP data, there is usually one pitch (positive label) in a frame while in MP data, there are usually multiple pitches in a frame.
In real settings, we do not know whether the music data is SP or MP in advance, so neither SPP or MPP algorithms do well in generic settings.

To address above challenges, we propose a highly accurate method for joint estimation of pitch, onset and offset (JEPOO).
Challenge 1 arises from three problems, (i) model design, (ii) data imbalance and (iii) multi-task weight allocation. 
To address problem (i), we design a model structure which has parameter sharing and feature fusion. 
Focal loss~\cite{lin2017focal} is a popular approach to problem (ii), and Pareto optimization~\cite{lin2019pareto} is a popular approach to problem (iii). 
However, we find that a direct application of focal loss or Pareto optimization separately yields very limited performance gain. 
Further, the gain of applying focal loss and Pareto optimization together is less than the sum of the gains of applying each technique separately. 
We believe this is because the weights obtained by one technique may conflict with those obtained by the other technique to some extent. Therefore, we propose a novel way to combine the two as follows. 
In focal loss, the $(1-\hat{y})^{\gamma}$ value is used to set the weight for each sample.
Since Pareto optimization produces the weights of the different tasks, we replace the  $(1-\hat{y})^{\gamma}$ value in focal loss by the task weight resulted from Pareto optimization.
The intuition is that the higher the weight of a task, the higher the weight of the samples in that task.
This way, we achieve much higher accuracy when using Pareto optimization together with focal loss, and we call the resulted loss as \textit{Pareto modulated loss} (PML).
Moreover, to avoid imbalance between the losses of different tasks, we impose a regularization on the weights of the losses of the three tasks, which we call \textit{loss weight regularization} (LWR).
The ultimate optimization method for JEPOO is PML with LWR.

Challenge 2 is caused by the different decision boundaries of SPP and MPP algorithms resulted from the positive/negative label imbalance. Our proposed PML has the effect of focal loss, which addresses data imbalance and enlarges the difference of the prediction values of positive and negative samples for both SP and MP data. 
Therefore, our method is robust w. r. t. different decision boundaries (Figure~\ref{fig:th}) and addresses Challenge 2.

Our contributions are summarized as follows: 
i) We propose JEPOO, a highly accurate method for joint estimation of pitch, onset and offset. 
We address the challenges of joint learning optimization and handling different types of data by novel model design and a new optimization technique named Pareto modulated loss with loss weight regularization. 
ii) This is the first work that can accurately handle both SP and MP music data or a mix of them. 
iii) A comprehensive experimental study on a wide range of real datasets shows that JEPOO significantly outperforms state-of-the-art methods by up to 10.6\%, 8.3\% and 10.3\% for the prediction of Pitch, Onset and Offset, respectively. 
Moreover, JEPOO's performance is robust for different types of datasets and instruments.

\section{Related Work}
\subsection{Pitch Prediction}
Pitch prediction, also termed pitch estimation, has been studied extensively~\cite{pitch}. 
Once we have extracted the pitches, then we can use them as features for MIR by recent recommendation algorithms such as~\cite{su2021detecting,wang2021combating} or they can be processed as time series using temporal databases~\cite{nmd}. 
Existing work on pitch estimation largely falls into two categories, SPP and MPP.

For SPP, traditional heuristic methods, such as ACF ~\cite{dubnowski1976real}, YIN ~\cite{de2002yin} and pYIN~\cite{mauch2014pyin}, employ a certain candidate-generating function to produce the pitch curve. Recently, some neural network based models have been proposed, such as CREPE ~\cite{kim2018crepe}, SPICE ~\cite{gfeller2020spice}. The accuracy of these
methods is limited because they do not utilize the timing
of the notes and hence cannot learn long-term sequential
patterns.

For MPP, there are mainly two types of methods, including frame-level transcription methods and note-level transcription methods\cite{benetos2018automatic}. The frame-level transcription methods, such as OAF \cite{hawthorne2017onsets}, ADSRNet\cite{kelz2019deep}, Non-Saturating GAN \cite{kim2019adversarial}, using CNN and LSTM to predict pitch results in each frame. While note-level transcription models, such as sequence-to-sequence \cite{hawthorne2021sequence} and MT3 \cite{gardner2021mt3} formulate the note as event to get the predictions using Transformer. But there is no research try to unify SPP and MPP algorithms as far as we know.

\subsection{Joint Learning in Melody Extraction}
Joint learning has been applied to MIR. For example, in \cite{choi2017transfer}, a transfer learning approach is used to solve classification and regression tasks in MIR simultaneously. In \cite{bittner2018multitask}, a multi-task deep learning model is used for melody, vocal and bass line estimation tasks. 
Besides, Hawthorne \emph{et al.} proposed OAF~\cite{hawthorne2017onsets} to jointly learn pitch, onset/offset together, by using onset/offset predictions to rectify pitch predictions. There are several follow-ups \cite{hawthorne2018enabling,kim2019adversarial,kelz2019deep}.
Unlike OAF, ADSRNet \cite{kelz2019deep} shares the bottom parameters and uses a strong temporal prior in the form of a handcrafted HMM to rectify pitch predictions. 
However, these joint models do not consider the balance between different tasks at all.

In order to balance different tasks in joint learning, there are some optimization methods for multi-task learning, such as GradNorm \cite{chen2018gradnorm}, DWA \cite{liu2019end}, DTP \cite{guo2018dynamic} and Pareto~\cite{lin2019pareto}. GradNorm and DWA make each task learn at a similar rate. DTP allows the model to give difficult task a bigger weight so as to dynamically prioritize difficult tasks during training. And Pareto determines the weight of each task through Pareto optimal solutions. However, none of them has been used in MIR as far as we know.

\section{Problem Formulation}\label{sec:task_form}
In this section, we formulate the problem of \textit{melody extraction}, for both SPP and MPP. 
In some studies, this is also called \textit{music transcription}.
The input is a Mel-spectrogram as shown in Figure~\ref{fig:melodyextraction}, which is a 2-dimensional matrix $\text{X}_{T\times F}$, where $T$ is the count of audio frames and $F$ is the number of frequency bins. Melody extraction coverts the Mel-spectrogram into a sequence of pitch values representing the f0 of the note; these are integer values typically in the range of $21$ to $108$. Thereby, melody extraction needs to predict the pitch value of each note, the starting frame of the note (onset) and the ending frame of the note (offset).
In the literature, these are treated as multi-label classification problems for each frame. Specifically, each frame $t$ of an audio is labeled by three 88-element one-hot arrays, $[\mathbf{y}^{\textit{p}},\mathbf{y}^{\textit{on}},\mathbf{y}^{\textit{off}}]$. The elements of $\mathbf{y}^{\textit{p}}$, $\mathbf{y}^{\textit{on}}$ and $\mathbf{y}^{\textit{off}}$ correspond to pitches, onsets and offsets, respectively.
Thus, the melody extraction task can be formally written as $\mathcal{F}: X_{T\times F} \to Y_{T\times 3 \times 88}$.

\section{Method}
To address the challenges discussed in Section~\ref{sec:intro}, we firstly design a model structure suitable for joint learning described in Section~\ref{sec:model}. Then we present our optimization technique PML with LWR in Section~\ref{sec:optimization}.

\subsection{Model Structure Design}\label{sec:model}
\begin{figure}[htbp]
\centerline{\includegraphics[width=0.8\columnwidth]{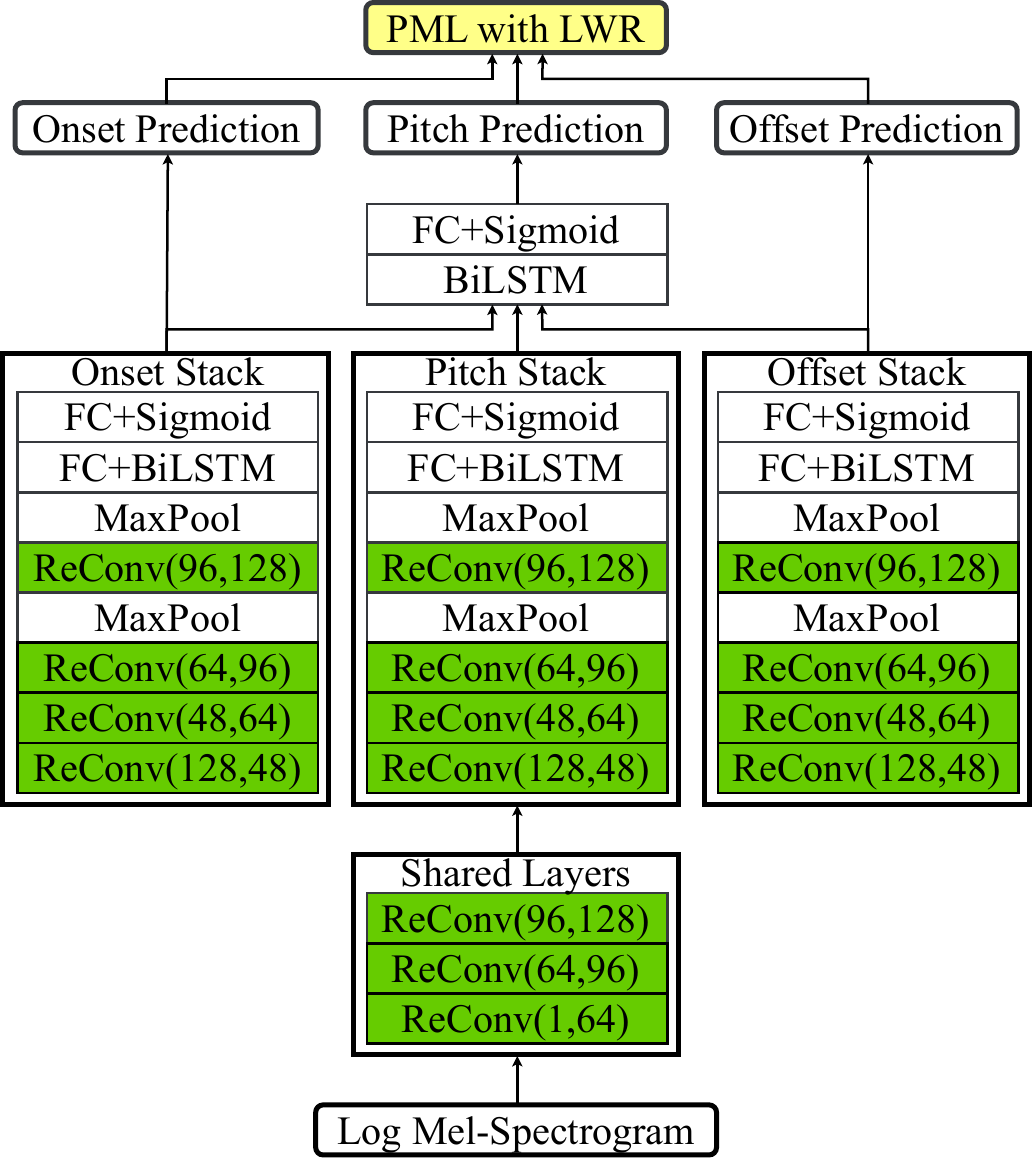}}
\caption{The overall structure of JEPOO.}
\label{fig:structure}
\end{figure}

The overall structure is illustrated in Figure~\ref{fig:structure}, 
which has three key mechanisms designed for the joint learning of pitch and onset/offset prediction: (i) Shared bottom layers, (ii) task-specific multi-label sequential predictors, and (iii) fusion of high level features, which are detailed below.

\begin{figure}[htbp]
\centerline{\includegraphics[width=0.5\columnwidth]{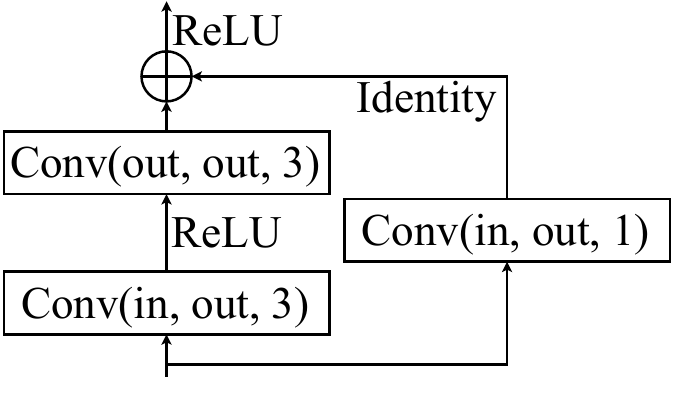}}
\caption{The details of residual convolution (ReConv) block.} 
\label{fig:Basic_Block}
\end{figure}

\textbf{Shared bottom layers.}
To capture common features of all sub-tasks, we stack several ReConv blocks as the shared bottom as shown in Figure~\ref{fig:Basic_Block}. 
A ReConv block contains two base convolutional layers and a skip-connection layer. The skip-connection layer is a convolutional layer with $1\times1$ kernel, while other kernels of base convolutional layers are $3\times3$. 
After element-wisely summing the output of skip-connection layer and last base convolutional layer, the result goes through a Relu function and becomes the output. 
By using ReConv block, the model can utilize multi-level features and become much deeper than conventional melody extraction models. 

\textbf{Task-specific multi-label sequential predictors.}
A note may contain hundreds of frames, so we design sequential predictors to utilize long-term features to achieve better performance.
Specifically, we design a task-specific sequential block, which consists of 4 ReConv blocks, a max pooling layer, a BiLSTM and a full connection layer with sigmoid. The input of task-specific multi-label sequential predictors is the output of Shared bottom layer. We call the parameters of different tasks as onset stack, pitch stack and offset stack. The output of these stacks is 88-elements vectors, whose elements are the predicted probability of corresponding labels. 

\textbf{Fusion of high level features.}
We use the predictions of onset and offset to help the pitch prediction in our model. 
Specifically, we concatenate the output of three stacks as the input of a BiLSTM layer along with full connection layer and sigmoid function to get pitch predictions. We do not use the pitch prediction to help onset/offset prediction due to the following reason. Onset/offset data only has positive labels in the starting/ending frames of a note, while pitch data has positive labels in almost all the frames of a note. The different distribution of pitch labels and onset/offset labels cause two impact. On the one hand, the positive labels of pitch between the starting and ending frames can be noise to Onset/Offset prediction. On the other hand, pitch prediction are inaccurate at boundaries of notes, and the onset/offset prediction can rectify the boundaries of note. Our experimental study validates the effectiveness of this design (see Appendix~\ref{sec:fusion_features}).

\subsection{Pareto Modulated Loss With Loss Weight Regularization (PML with LWR)}\label{sec:optimization}
As discussed in Section~\ref{sec:intro}, to address the problems of data imbalance and multi-task weight allocation, we propose to use a combination of focal loss and Pareto optimization. Next, we firstly describe a naive combination of Pareto optimization and focal loss below, but it only yields limited performance improvement. We then present Pareto Modulated Loss (PML) with Loss Weight Regularization (LWR), which works much better.

\textbf{Naive Combination of Pareto Optimization and Focal Loss (Naive Optimization).}
Focal loss is a popular technique to alleviate the category imbalance within a sub-task by setting hyper-parameter $\alpha$ and shift the decision boundaries. Pareto optimization is popular technique to allocate the weight of sub-tasks. A naive way to combine these two techniques is to multiple the Focal loss with the task weight produced by Pareto optimization as follows, and we call it the naive optimization.
\begin{equation} 
\mathcal{L}_{task} = -\sum_{i}^{n} \omega^{i}\alpha^{i} y^{i} (1-\hat{y}^{i})^{\gamma^{i}} \log{\hat{y}^{i}}
\end{equation}
where  $\omega^{i}$ is a Pareto optimal solution for \emph{i}th sub-task and $y^{i}$ is the label of \emph{i}th sub-task, $n$ is the total number of sub-tasks.

Our empirical evaluation in section \ref{sec:ablation} shows that the naive optimization addresses data imbalance and weight allocation to some extent, but the improvement is limited. The reason may be as follows. Firstly, focal loss needs to search two hyper-parameters for each sub-task, resulting in high cost of grid search. As we mentioned above, data of different tasks varies hugely, so it is hard to determine the scale and precision of data related hyper-parameter $\gamma$ for each sub-task and results in sub-optimal hyper-parameters. What's worse, as the number of tasks grows, grid search space grows exponentially. Secondly, Pareto optimal solution may be imbalance between tasks, resulting in some tasks hard to optimize. The model may need to be trained many times before making all the tasks are optimized.

\textbf{PML with LWR.}
To solve the above problems, we propose a novel Pareto Modulated Loss (PML) by integrating Pareto optimization and focal loss to reduce the high training cost of focal loss. In addition, we design a loss weight regularization method to avoid the imbalance between the loss weights of the Pareto optimal solution.

PML is based on our observation that the difficulty of a task reflects the difficulty of its own data. Thus, we try to replace the item $(1-\hat{y})^{\gamma}$ of focal loss by the task weight produced by Pareto optimization. This way, data in different batches have different weights, and the grid search space becomes at 50\% less than the focal loss.

After getting the Pareto optimal weights $[\omega^{1},...,\omega^{n}]$ of different tasks' loss, 
we use a MLP layer to get the final weights. i.e. $[\omega^{1}_{PML},..,\omega^{n}_{PML}] =softmax(\mathbf{W}[\omega^{1},...,\omega^{n}]+\mathbf{b})$.
The formal form of PML is written as follows:
\begin{equation}
    \mathcal{L}_{task} = -\sum_{i}^{n} \omega^{i}_{PML}\alpha^{i}y^{i}\log{\hat{y}^{i}}
\end{equation} 
PML obtains more discriminative ability than Pareto optimization without introducing any hyper-parameter. Moreover, PML has the effect of focal loss, which enlarges the difference of the prediction values of positive and negative samples.

To avoid imbalance of loss weights, we design a simple yet efficient Loss Weight Regularization (LWR). The LWR item is calculated as follows:
\begin{equation}
    \mathcal{L}_{re} = \sum_{i}^{n} ||n\omega^{i}_{PML} -1||_{p}
\end{equation} $\omega_{PML}$ represents the loss weight of different sub-tasks. Because $\sum_{i}^{n}\omega^{i}_{PML} = 1$, the LWR gives penalty to the weight that is far away from the average.

Then final PML with LWR loss is defined as follows:
\begin{equation}
    \mathcal{L}_{total} = \mathcal{L}_{task} + \lambda \mathcal{L}_{re}
\end{equation}
Compared with PML, 
PML with LWR only introduces one more hyper-parameter than Pareto optimization, the weight of LWR item $\lambda$. The grid search cost is much less than the native optimization.

\section{Experimental Setup}

\begin{table*}[htp]
\centering
\resizebox{0.9\textwidth}{!}{
\begin{tabular}{c|ccc|ccc|ccc}
\hline
\multirow{2}{*}{Methods}& \multicolumn{3}{c|}{F1(\%) on MDB-stem-synth(SP)} & \multicolumn{3}{c|}{F1(\%) on MAPS(SP\&MP)} & \multicolumn{3}{c}{F1(\%) on MAESTRO(MP)} \\ 
 & Pitch & Onset & Onset\&Offset & Pitch & Onset & Onset\&Offset & Pitch & Onset & Onset\&Offset \\\hline
PYIN*~\cite{mauch2014pyin}&79.6   &56.5   &56.4 &12.5  &28.4  &28.1  &10.5 &25.0  &24.4          \\ 
CREPE*~\cite{kim2018crepe}& 90.6  & 78.5  & 78.5          & 26.0  & 41.3  & 40.8          & 21.5  & 41.3  & 40.2          \\ 
OAF-retrain~\cite{hawthorne2017onsets} &95.3  &90.9 &89.8 &79.7 &81.9 & 61.7 &89.7 &94.1  &79.6    \\
OAF* ~\cite{hawthorne2017onsets} &65.5  &38.2 &26.5 &71.7 &80.8 &40.8 &90.2 &95.3  &80.5    \\
ADSRNet$^\dag$~\cite{kelz2019deep}& ——  & ——  & ——     & 77.2     & 81.4     & 56.1    & ——        & ——        & ——             \\ 
Non-Saturating GAN$^\dag$ ~\cite{kim2019adversarial} &—— &—— &—— &—— &—— &—— &91.4  &95.6   &81.3 \\
KJN$^\dag$~\cite{kwon2020polyphonic}    &—— &—— &—— &—— &—— &—— &83.8 &94.7  &79.4 \\ 
sequence-to-sequence*~\cite{hawthorne2021sequence}       &20.0 &29.8 &22.7 &47.1   &75.7  &35.4     & 66.0 & 96.0 & 83.5             \\ 
MT3*~\cite{gardner2021mt3}        &12.0 &4.8  &2.3 &74.4       &80.7       &51.6    & 86.0    & 95.0   & 80.0               \\ \hline
JEPOO & \textbf{97.1}  & \textbf{96.0}  & \textbf{95.6}          & \textbf{81.6}  & \textbf{84.2}  & \textbf{65.6}  &\textbf{93.0}  & \textbf{96.5}  & \textbf{84.0} \\ \hline
\end{tabular} 
}
\caption{Performance comparison on both SP and MP datasets in terms of F1 score. * means we reproduce the results using authors' open source checkpoints. $^\dag$ represents we copy the results from original papers. OAF-retrain represents retraining OAF on three open datasets.}
\label{table:main_results}
\end{table*}

\textbf{Datasets.} To compare with previous pitch prediction methods on SP and MP data, we use three real datasets MDB-stem-synth ~\cite{salamon2017analysis}, MAPS ~\cite{emiya2009multipitch} and MAESTRO-V1.0.0 ~\cite{hawthorne2018enabling}. 
\textit{MDB-stem-synth} is a SP dataset. It contains 230 resynthesized monophonic music files spanning 25 musical instruments corresponding perfect f0 annotation. 
\textit{MAPS} has a very small percentage of SP data and a majority of MP data. It contains 270 raw audio recordings of piano music and corresponding MIDI-annotated piano recordings. 
\textit{MAESTRO-V1.0.0} is a MP dataset larger than MAPS. It contains 172.3 hours of paired audio and MIDI recordings from ten years of International Piano-e-Competition. 

\noindent\textbf{Evaluation Metrics.} Following MT3~\cite{hawthorne2017onsets} and CREPE~\cite{kim2018crepe}, we use 4 metrics to evaluate our model. 
These metrics are computed by mir\_eval \cite{raffel2014mir_eval}. 
The details are described as follows:
\textit{The F1 score of pitch prediction (Pitch)} uses a binary measure of whether the prediction of a frame and the ground truth matches. 
Each second will be divided into a fixed number of frames, and the sequence of notes is represented as a binary matrix of size [frames × 88], which indicates the presence or absence of an active note at a given pitch and time.
\textit{Note with onset (Onset)} considers a prediction to be correct if it has the same pitch and is within ±50ms of a reference onset.
\textit{Note with onset and offset (Onset\&Offset)}. In addition to matching onsets and pitches as above, this metric requires the note to also end in this frame (offset). 
\textit{Voicing false alarm rate (VFA)} computes the proportion of non-melody frames in the ground truth but are mistakenly predicted as melody frames. 

\section{Experimental Results}\label{sec:experiments}
Firstly, we compare JEPOO with SOTA methods on both real datasets and synthetic datasets of mixed SP and MP data. We then perform an ablation study to understand the effectiveness of the techniques we propose. Finally, we investigate the robustness of JEPOO on different types of datasets and instruments. The implementation details and comparison systems are provided in Appendix~\ref{sec:im_details} and~\ref{sec:comparison} respectively.

\subsection{Main Results}
\noindent\textbf{Experiment on real datasets.} Table~\ref{table:main_results} shows the results of comparing JEPOO with SOTA methods on the three aforementioned real datasets. We observe that JEPOO outperforms all the other methods on the datasets and all the tasks (Pitch, Onset, Onset\&Offset) consistently. JEPOO outperforms the naive joint learning method OAF at all metrics by up to 30\%, 58\%, 69\% in pitch, onset and offset prediction, respectively. This confirms that naive joint learning is far from optimal and our methods are necessary. JEPOO also outperforms SPP methods (CREPE, PYIN) on SP data, and MPP methods (all the rest) on MP data, respectively. Even we retrain OFA on SP data, and get the improved OAF-retrain method, JEPOO still outperforms OAF-retrain significantly on all tasks. These results show the effectiveness of our model design and optimization techniques. 
Further analysis of different optimization techniques are provided in Section~\ref{sec:ablation}.

Moreover, it should be noted that existing methods do not perform well on SP and MP data, simultaneously. As shown in Table \ref{table:main_results}, SPP models (CREPE and PYIN) perform poorly on MP datasets. MPP models (OAF, MT3 and sequence-to-sequence) perform poorly on SP datasets.
\begin{figure*}[htbp]
\centerline{\includegraphics[width=0.8\textwidth]{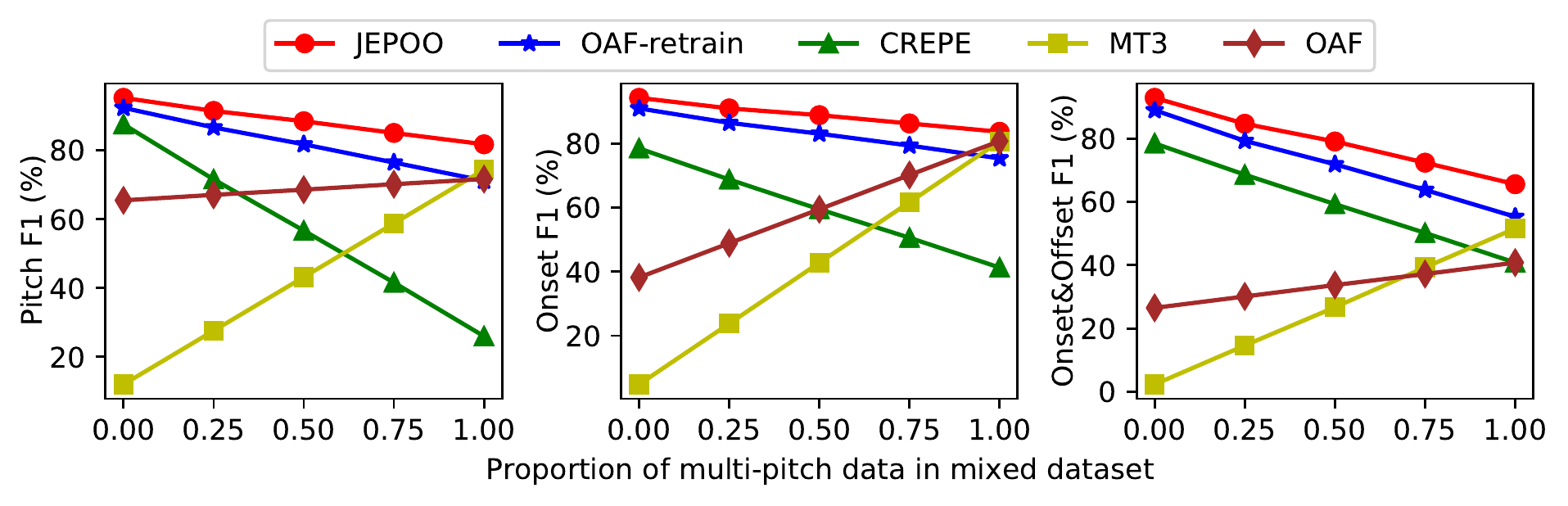}}
\caption{Performance on synthetic test datasets with different proportion of multi-pitch data. The results of CREPE, MT3 and OAF are reproduced by using authors' open source checkpoints. OAF-retrain represents retraining OAF on synthetic train dataset.}
\label{fig:rate_of_MAPS}
\end{figure*}

\noindent\textbf{Experiment on synthetic dataset mixing SP and MP data.}
In real application scenarios, melody extraction models need to handle both SP and MP music data, because we are unable to know the data type in advance. Unfortunately, the current datasets are not suitable to evaluate such ability. Though MAPS is created to evaluate both SPP and MPP, the amount of different data are unbalance. Multi-pitch frames are 6 times as many as single-pitch frames. Worse still, the proportion of single-pitch frames in real audios is unknown. 

To create a dataset that can evaluate the ability of JEPOO on handling both SP and MP data, we take two steps: \textit{Firstly}, we mix the train dataset of MAPS and MDB-stem-synth to create synthetic train dataset.
\textit{Secondly}, to simulate various situations of real audio, we create multiple test datasets with different proportions of multi-pitch data by adding different amount of MDB-stem-synth test data into MAPS test dataset. The proportion ranges from 0 to 1 with the step of 0.25.

We retrain JEPOO and OAF on the synthetic train dataset. Figure~\ref{fig:rate_of_MAPS} shows the evaluation results on different synthetic test datasets. In this figure, JEPOO outperforms all the other methods at any proportion of MP data significantly. JEPOO outperforms SOTA method by up to 10.6\%, 8.3\% and 10.3\% at pitch, onset and offset prediction, respectively. By contrast, current models only work well on one type of data. For example, the SPP model CREPE decreases fastest as the raising of the proportion of MP data. MT3 performs worst when SP data is in majority. OAF-retrain gets second best results, but the gap of JEPOO and OAF-retrain increases with increasing of the proportion of MP data. These results indicate the ability of JEPOO to handle SP and MP data simultaneously.

Through comparing with Table \ref{table:main_results}, we find that OAF trained on synthetic dataset get worse performance than the model trained on SP and MP dataset respectively, decreasing at pitch prediction by 2.9\% on MDB-stem-synth test dataset and 8.5\% on MAPS test dataset. While JEPOO trained on synthetic dataset improves 0.2\% on MAPS test dataset. This result indicates that simply training model on synthetic data cannot improve model performance. Experiments on synthetic dataset of mixing SP and MP data show the ability of JEPOO to handle both SP and MP music data, which has great practical value in MIR systems.

\subsection{Ablation Study} \label{sec:ablation}
In this experiment, we not only evaluate different optimization techniques, 
but also compare JEPOO with naive joint learning, single task training on synthetic datasets. We conducted ablation studies of each component of JEPOO. 
More ablation studies about optimization techniques, sequential predictor, model structure and feature fusion are given in Appendix~\ref{sec:ablation_optim},~\ref{sec:bilstm_transformer},~\ref{sec:cnn_layers} and~\ref{sec:fusion_features}, respectively. 

From Table~\ref{table:ablation study},
\begin{table}[htbp]
\centering
\resizebox{\columnwidth}{!}{
\centering
\begin{tabular}{c|ccc}
\hline
\multirow{2}{*}{Methods}& \multicolumn{3}{c}{F1(\%) on MAPS+MDB-stem-synth }            \\ 
                      & Pitch & Onset & Onset\&Offset \\ \hline
JEPOO               & \textbf{87.6}  & \textbf{88.3}  & \textbf{77.4} \\ 
JEPOO with naive optimization                  & 87.2  & 87.9  & 76.8    \\
JEPOO with only FL                & 86.9  & 87.5  & 76.4    \\
JEPOO with only Pareto            & 86.4  & 87.9  & 76.1    \\ \hline
Naive joint learning of pitch, onset and offset & 86.0  & 87.2 & 74.4  \\ 
Naive joint learning of pitch and onset& 85.5  & 87.4 & 74.3  \\ \hline
Single model of pitch (SMP)& 86.7  & 80.1 & 68.3  \\ \hline
\end{tabular}
}
\caption{Ablation study on different optimization techniques in terms of F1 score. The test dataset is 1:1 mixed SP and MP data.}
\label{table:ablation study}
\end{table}
we observe that naive joint learning does not improve all sub-tasks, though pitch, onset and offset are highly related. For example, the naive joint learning of pitch, onset and offset decreases 0.7\% at Pitch than SMP. While JEPOO with naive optimization and JEPOO both outperform naive joint model at all metrics. Although JEPOO with naive optimization performs better than JEPOO with only FL and JEPOO with only Pareto, but the improvement is only 1.2\% at Pitch, which is less than the sum of separated improvement of focal loss (0.4\%) and the Pareto optimization (0.9\%). In addition, JEPOO with naive optimization achieves the same performance at Onset than JEPOO with only Pareto. JEPOO improves 1.6\% at Pitch, 3.0\% at Onset\&Offset than naive joint model. Besides, the training cost of naive optimization is about four times as PML with LWR. Above results show that PML with LWR can better balance different sub-tasks and data in a low training cost than naive optimization.

\subsection{Robustness of JEPOO}
\textbf{Experiment with different instruments.}
It is obviously that the performance of melody extraction varies on different instruments. To investigate the impact of instruments on our method, we deeper evaluate JEPOO on the multi-instrument dataset MDB-stem-synth along with OAF and CREPE. The reason that we only compare with these baselines is that OAF-retrain achieves second best results on MDB-stem-synth and CREPE is designed for this dataset.
In this experiment, we use stratified sampling to split the train and test dataset, ensuring that the distribution of instruments is consistent between two datasets. 
By this way, the train and test datasets contain 25 instruments, such as bass, violin, flute and singing voice etc. We retrain JEPOO, OAF on the new multi-instrument dataset and evaluate these models on the new test dataset. 

\begin{figure*}[htbp]
\centerline{\includegraphics[width=0.9\textwidth]{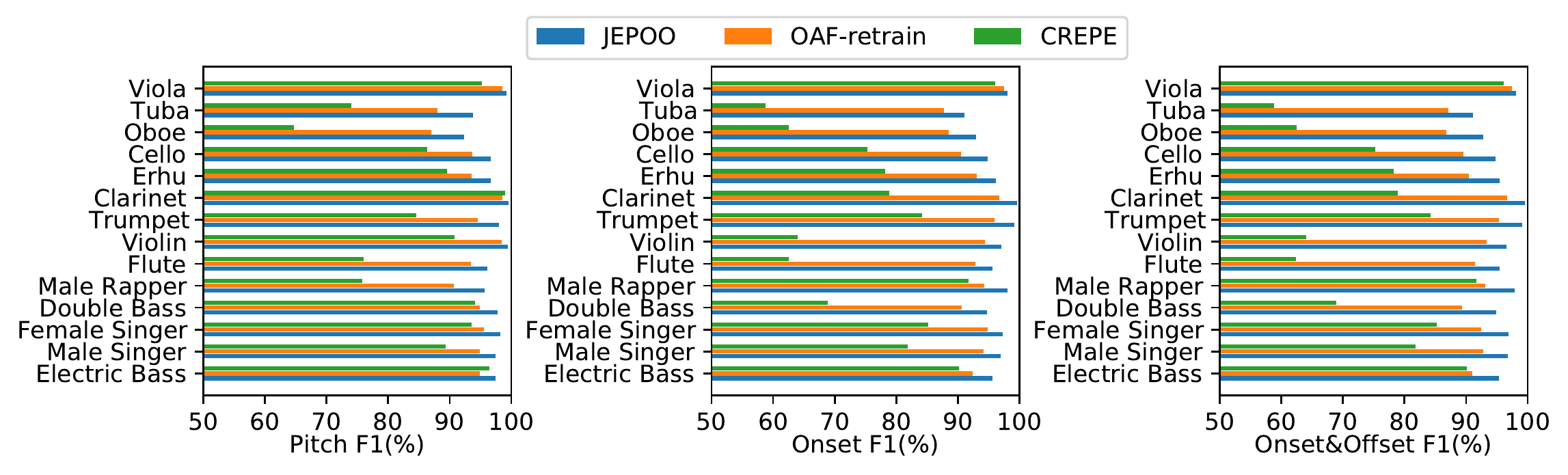}}
\caption{The performance of different models with different instruments. OAF-retrain represents retraining OAF.}
\label{fig:instruments}
\end{figure*}

The results on various instruments are displayed in Figure~\ref{fig:instruments}. 
From this figure, 
there are following conclusions:
\textbf{i)} JEPOO consistently outperforms OAF-retrain and CREPE at all metrics on all instruments. This result shows better discrimination ability of our method on different instruments than SOTA methods.
\textbf{ii)} Our method has excellent generality on different instruments. From Figure~\ref{fig:instruments}, we can see that JEPOO achieves more stable performance than OAF-retrain and CREPE. The F1 score variances of JEPOO are only 4.0, 5.1 and 5.0 at Pitch, Onset and Onset\&Offset, respectively. While the variances of OAF-retrain are 11.5, 8.0 and 9.6, and the variances of CREPE are 95.2, 135.1 and 135.1. The high variance of CREPE is because the checkpoint is trained on unbalanced instrumental datasets. 
The above conclusions demonstrate the high accuracy and robustness of JEPOO on multiple instruments. 


\noindent\textbf{Robustness on SP and MP datasets.}
To evaluate the robustness of JEPOO on handling SP and MP data, we train JEPOO and OAF on the synthetic train dataset and evaluate them with different positive thresholds on SP and MP dataset, respectively. There is a positive prediction when predicted probability is larger than the positive threshold. We test positive thresholds from 0.1 to 0.9 with step 0.1. The left of Figure~\ref{fig:th} reports the result on SP dataset MDB-stem-synth and the right of Figure~\ref{fig:th} reports the result on MP dataset MAPS.
According to Figure~\ref{fig:th}, we can draw following conclusions:
\begin{figure}[htbp]
\centerline{\includegraphics[width=0.85\columnwidth,height=1.05\columnwidth]{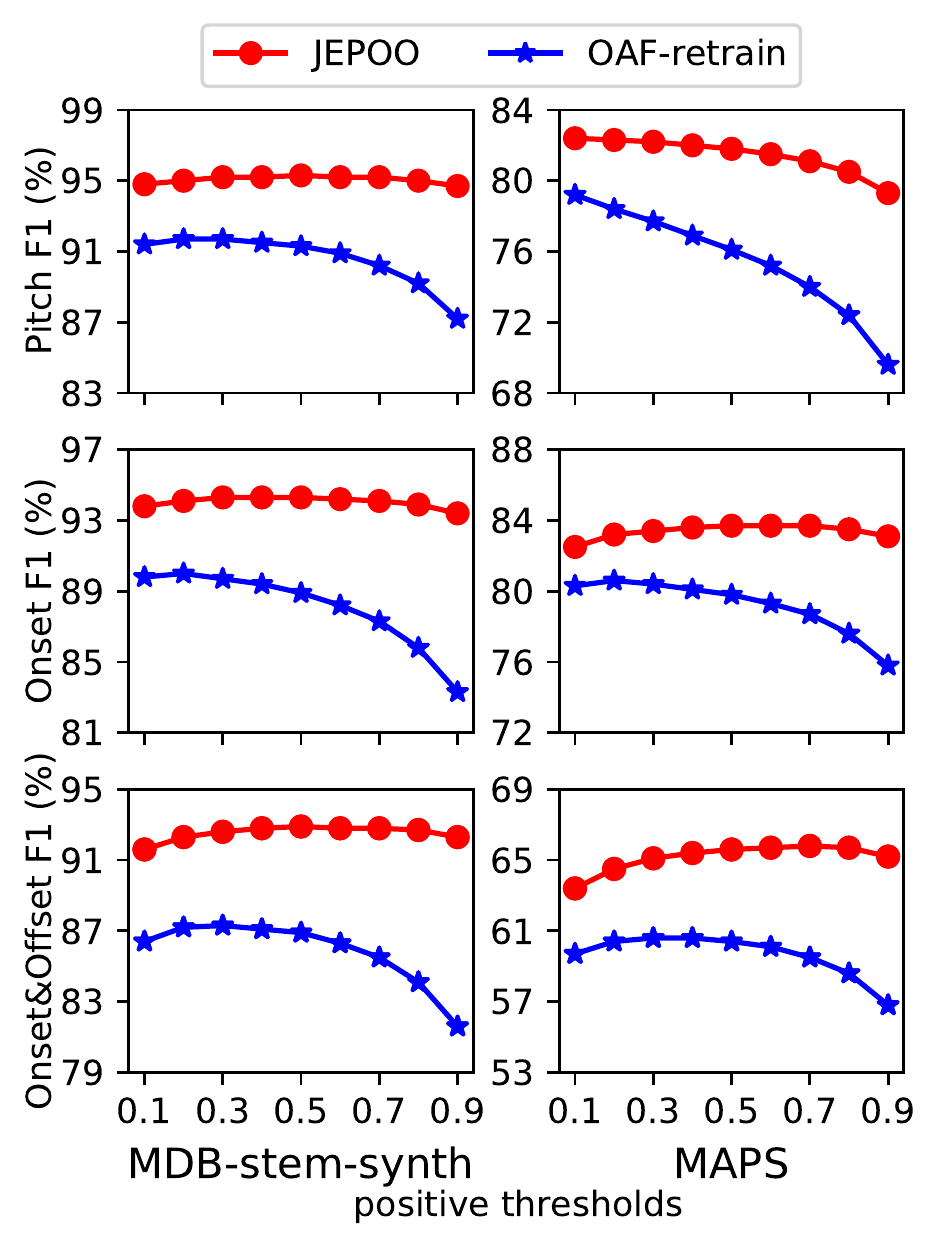}}
\caption{Comparison of different thresholds on MAPS (MP) and MDB-stem-synth(SP).}
\label{fig:th}
\end{figure}

Firstly, JEPOO outperforms OAF-retrain on MDB-stem-synth and MAPS at all thresholds. Specifically, on MDB-stem-synth dataset, OAF-retrain achieves 91.7\%, 90.0\% and 87.2\% at Pitch, Onset and Onset\&Offset with the best threshold. JEPOO achieves 95.3\%, 94.3\% and 92.8\% at Pitch, Onset and Onset\&Offset when the best threshold is 0.5. There is a similar result on MAPS dataset. These results indicate that JEPOO can separate positive and negative samples clearly, making it robust against noise.

Secondly, JEPOO is more robust on both SP and MP data. The best threshold of JEPOO is near 0.5 on both SP and MP dataset, while the best threshold of OAF-retrain is just 0.2. When the threshold grows from 0.1 to 0.9, JEPOO decreases less than 0.7\% at Pitch, 1.0\% at Onset and 1.4\% at Onset\&Offset on SP dataset, and less than 3.1\%, 1.2\% and 2.4\% on MP dataset. However, OAF-retrain decreases significantly when the threshold grows from 0.1 to 0.9, up to 9.6\% at Pitch on MP dataset. Above results show the robustness of JEPOO on handing SP and MP datasets.

\noindent\textbf{Non-Melody Frames Robustness.} To evaluate robustness on non-melody frames, we use Voicing false alarm rate (VFA) to evaluate different models and report the results in Table \ref{tab:case}. 
The smaller at VFA, the better performance of the model.
In this experiment, we add white noise into clean audios to simulate real audios. We use Signal-to-noise ratio (SNR) to measure noise.
The higher of SNR means the less of noise. Thus, the SNR INF in Table~\ref{tab:case} represents raw clean audio.

JEPOO hardly predict pitch at non-melody frames on clean audios. In Table ~\ref{tab:case}, JEPOO achieves only 0.2\% at VFA, while CREPE is 48 times higher. 
\begin{table}[htbp]
\centering
\resizebox{0.9\columnwidth}{!}{
\begin{tabular}{c|cc}
\hline
\multirow{2}{*}{Methods}& \multicolumn{2}{c}{VFA(\%) on MDB-stem-synth}            \\ 
                      & SNR INF &  SNR 50              \\\hline 
CREPE~\cite{kim2018crepe}                    & 9.8 &28.0         \\ \hline
OAF-retrain~\cite{hawthorne2017onsets}                      & 0.5  &11.8         \\ \hline
JEPOO       & \textbf{0.2} &\textbf{3.6} \\ \hline
\end{tabular}
}
\caption{Comparison of different models at non-melody frames in terms of VFA score.}
\label{tab:case}
\end{table}
Besides, when SNR is down to 50, the VFA of our model only increase to 3.6\%, while OAF-retrain is up to 11.8\% and CREPE is up to 28\%. Although OAF-retrain only achieves 0.5\% at VFA when no noise, but we find OAF-retrain predicts a pitch when the probability is higher than 0.03. This relative low threshold makes OAF is influenced by noise easily. As our model, the frame will be predicted positive only when the output probability is greater than 0.5. The above results show that JEPOO is more robust on non-melody frames and can predict correctly on almost all non-melody frames, since our model can separate positive and negative samples more clearly.

\begin{figure}[htbp]
\centerline{\includegraphics[width=\columnwidth,height=0.45\columnwidth]{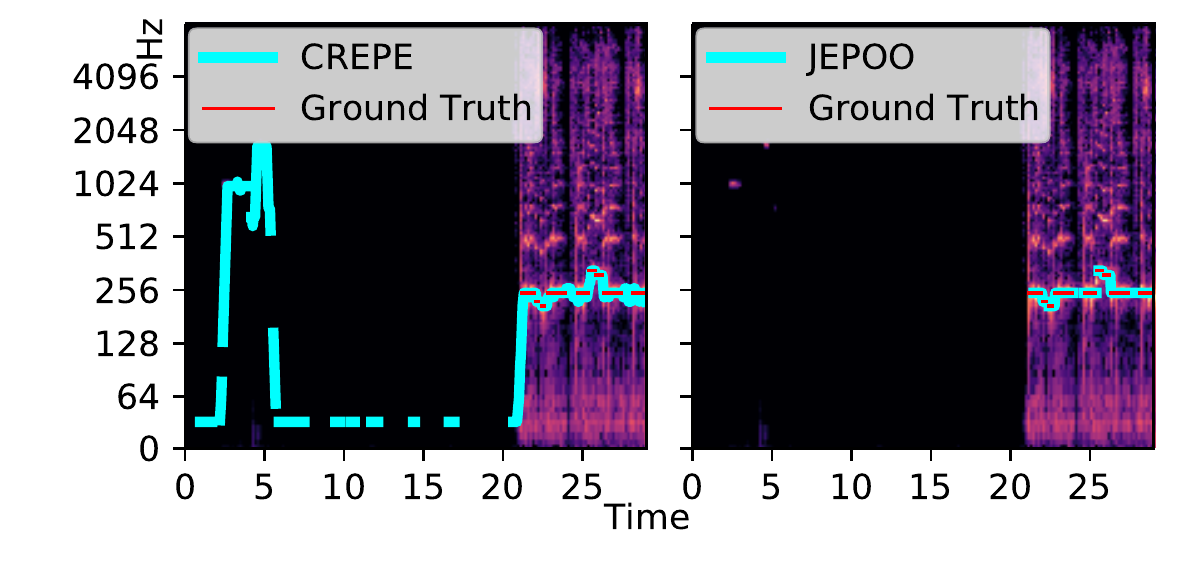}}
\caption{Case study of Robustness. The left is the predictions of CREPE, and the right is the predictions of JEPOO.}
\label{fig:case}
\end{figure}

\noindent\textbf{Case study of Robustness.} To intuitively understand the performance at non-melody frames, we visualize the result of a case to compare JEPOO and CREPE in Figure~\ref{fig:case}. 
The test data is the clip from 12s to 41s of 403th song in the MIR\_ST500~\cite{mirst500_2021}. The clean vocal is obtained by ResUNetDecouple+~\cite{kong2021decoupling}. 
From this figure, we observe that our method predicts no pitch at non-melody frames, while CREPE predicts high pitch value at non-melody frames. This is because CREPE can be influenced by small noise (top left on mel-spectrogram) easily. In addition, the predictions of JEPOO coincide more closely with the ground truth, the red line in figure, than CREPE.  Above results indicate the robustness of JEPOO in another aspect.

\section{Conclusion}
In this paper, we propose JEPOO, a highly accurate method for joint estimation of pitch, onset and offset by proposing novel model design and a new optimization technique named Pareto modulated loss with loss weight regularization. JEPOO significantly outperforms state-of-the-art methods by up to 10.6\%, 8.3\% and 10.3\% for the prediction of Pitch, Onset and Offset, respectively, and JEPOO's performance is robust for different types of datasets and instruments.

\section*{Acknowledgments}
This work is supported in part by the National Science Foundation of China under the grant 62272466, and Public Computing Cloud, Renmin University of China.

\bibliographystyle{named}
\bibliography{ijcai23}
\clearpage
\appendix
\section{Appendix}\label{sec:appendix}
\subsection{Implementation Details} \label{sec:im_details}
The raw audio is sampled at 16 kHz and then transformed into Mel-spectrograms with log amplitude which has 229 mel bins. The hop length of Mel-spectrograms is 512 and the Hann window size is 2048. And we cut the frequency between 30Hz and 8000Hz to extract the Mel-spectrograms. Above audio processing uses librosa \cite{mcfee2015librosa}, which is the same as OAF\cite{hawthorne2017onsets}.

The input and output channels of all ReConv blocks are marked in Figure~\ref{fig:structure}, and all kernel sizes are introduced in Figure~\ref{fig:Basic_Block}. Kernel size of MaxPool is (1, 2). The dimension of input and output of BiLSTM are both 768. 
We set the batch size as 16 and use the Adam optimizer~\cite{kingma2014adam}. Each training audio is randomly selected from original audio with same duration, 12.8s. The learning rate is initialized as 0.0005 and reduced by 0.98 of the previous learning rate every 10000 iterations. 
We use the code released by~\cite{lin2019pareto} to implement Pareto optimization. 
The initial weights of different sub-tasks are all 1, and we search new weights every 10 iterations. In loss weight regularization, we use $L_2$ distance, and the weight $\lambda$ is 0.04.

We use the same way as CREPE\cite{kim2018crepe} to split MDB-stem-synth into train and test datasets. 
The split way of MAPS is the same as OAF \cite{hawthorne2017onsets}. The split way of MAESTRO are introduced in \cite{hawthorne2018enabling}. We process MIDI files of MAPS and MAESTRO using the same method as \cite{hawthorne2017onsets}. The raw labels in time, frequency are transformed to note onset time, note offset time and midi numbers similar to MAPS and MAESTRO, when processing the MDB-stem-synth dataset.

\subsection{Comparison System}\label{sec:comparison}
We compare JEPOO with some recently proposed melody extraction methods. 
CREPE ~\cite{kim2018crepe} and pYIN \cite{mauch2014pyin} are the SPP methods and have open source checkpoints. We do not compare our model with SPICE ~\cite{gfeller2020spice}, because SPICE does not open code and use different metrics.

OAF~\cite{hawthorne2017onsets} is the first joint model and it has been designed for multi-pitch estimation. Its source code is publicly available and it has been trained on MP datasets. As discussed in Section~\ref{sec:intro}, we may retrain it on SP data or mixed SP and MP data to get better performance, so we have retrained OAF on the various datasets respectively for each experiment and refer to this retrained version as \textit{OAF-retrain}.
The results of ADSRNet \cite{kelz2019deep}, Non-Saturating GAN~\cite{kim2019adversarial} and KJN~\cite{kwon2020polyphonic} are directly obtained from their original papers, because they do not make their code available.

MT3~\cite{gardner2021mt3} and sequence-to-sequence \cite{hawthorne2021sequence} are Transformer-based multi-pitch estimation models. The results of both models are reproduced by using authors’ open source checkpoints. We do not compare our model with SpecTNT ~\cite{lu2021spectnt}, because it does not open code and is evaluated on different datasets.

\subsection{Ablation Study of Optimization Techniques}\label{sec:ablation_optim}
\begin{table}[htbp]
\centering
\resizebox{\columnwidth}{!}{
\centering
\begin{tabular}{c|ccc}
\hline
\multirow{2}{*}{Methods}& \multicolumn{3}{c}{F1(\%) on MAPS}            \\ 
                      & Pitch & Onset & Onset\&Offset \\ \hline
JEPOO               & \textbf{81.8}  & \textbf{83.7}  & \textbf{65.6} \\ 
JEPOO with naive optimization                  & 81.5  & 83.1  & 64.9    \\
JEPOO with only FL                & 80.5  & 82.6  & 64.1    \\
JEPOO with only Pareto            & 80.8  & 82.2  & 63.9    \\ \hline
Naive joint learning of pitch, onset and offset & 79.8  & 82.1 & 63.5  \\ 
Naive joint learning of pitch and onset& 79.6  & 82.8 & 62.4  \\ \hline
Single model of pitch (SMP)& 80.2  & 75.1 & 58.6  \\ \hline
\end{tabular}
}
\caption{Ablation study on different optimization techniques in terms of F1 score. The test dataset is MAPS (MP) dataset.}
\label{table:ablation_maps}
\end{table}
\noindent In this section, we report more results of ablation studies for optimization techniques. Experimental settings are the same as those in Section~\ref{sec:ablation}.
We evaluate different optimization techniques on MAPS and MDB-stem-synth datasets, and the results are shown in Table~\ref{table:ablation_maps} and Table~\ref{table:ablation_mdb} respectively.
\begin{table}[htbp]
\centering
\resizebox{\columnwidth}{!}{
\centering
\begin{tabular}{c|ccc}
\hline
\multirow{2}{*}{Methods}& \multicolumn{3}{c}{F1(\%) on MDB-stem-synth}            \\ 
                      & Pitch & Onset & Onset\&Offset \\ \hline
JEPOO               & \textbf{95.3}  & \textbf{94.3}  & \textbf{92.9} \\ 
JEPOO with naive optimization                  & 94.9  & 94.1  & 92.5    \\
JEPOO with only FL                & 94.3  & 92.5  & 92.1    \\
JEPOO with only Pareto            & 94.1  & 91.4  & 92.0    \\ \hline
Naive joint learning of pitch, onset and offset & 93.8  & 90.4 & 91.6  \\ 
Naive joint learning of pitch and onset& 92.9  & 90.1 & 87.4  \\ \hline
Single model of pitch (SMP)& 94.2  & 82.5 & 80.8  \\ \hline
\end{tabular}
}
\caption{Ablation study on different optimization techniques in terms of F1 score. The test dataset is MDB-stem-synth (SP) dataset.}
\label{table:ablation_mdb}
\end{table}

\subsection{Comparison of BiLSTM and Transformer}\label{sec:bilstm_transformer}
In this section, we compare the effect of using BiLSTM or using Transformer in Table \ref{table:BT}. 
\begin{table*}[]
\centering
\resizebox{0.7\textwidth}{!}{
\centering
\begin{tabular}{c|ccc|ccc}
\hline
\multirow{3}{*}{(BiLSTM or Transformer, \#shared, \#unshared)}& \multicolumn{6}{c}{F1(\%) on MAPS+MDB-stem-synth}            \\ \cline{2-7}
                    &\multicolumn{3}{c|}{Base} &\multicolumn{3}{c}{PML+LWR} \\ \cline{2-7}
                    & Pitch & Onset & Onset\&Offset & Pitch & Onset & Onset\&Offset\\ \hline
OAF-retrain     & 82.7          & 83.8          & 71.9       &82.7         &83.8    &71.9  \\ \hline 
(BiLSTM, 6, 4)        & 86.8           & 87.6          & 75.4       &87.2          &88.0    &76.6       \\ \hline
(1 Transformer, 6, 4) & 85.0          & 86.9         & 72.5       &85.0         &87.0    &73.0       \\ 
(2 Transformer, 6, 4) & 86.5           & 87.9          & 75.8       &86.4          &88.2    &75.6       \\ 
(3 Transformer, 6, 4) & 86.8           & 88.1          & 76.2       &86.5          &88.1    &75.8       \\ 
(4 Transformer, 6, 4) & 86.5           & 87.7         & 75.7       &86.5          &87.5    &75.8       \\ \hline
(1 Transformer, 6, 8) & 85.5          & 87.7        & 73.7       &85.7          &87.7    &74.9       \\ 
(2 Transformer, 6, 8) & 86.1           & 88.0         & 75.5       &86.3          &88.5    &75.9       \\ 
(3 Transformer, 6, 8) & 86.9           & \textbf{88.1}          & 76.6      &86.9         &\textbf{88.6}    &76.8       \\ 
(4 Transformer, 6, 8) & 86.6           & 87.7         & 76.1       &86.9         &88.2    &76.4       \\ \hline
JEPOO (BiLSTM, 6, 8)       & \textbf{87.2}  & 87.9  & \textbf{76.8} &\textbf{87.6}         &88.3    &\textbf{77.4}       \\ \hline
\end{tabular}
}
\caption{The comparison between using BiLSTM or using different layers of Transformer in terms of F1 score. \#shared means the number of convolutional layers in shared layers, \#unshared means the number of convolutional layers for specific sub-tasks. The test dataset is 1:1 mixed SP and MP data.}
\label{table:BT}
\end{table*}
We use Transformer with different layers to replace every BiLSTM in Figure~\ref{fig:structure}. Besides, we also report the performance with different optimization techniques that we proposed. From this Table, we find that models with BiLSTM achieves better performance at two of the three metrics, Pitch and Onset\&Offset. Based on this result, we adopt BiLSTM in our model design, rather than Transformer. In addition, 70\% of models using PML with LWR have better performance than the one using the naive optimization technique. This result indicates the generality of PML with LWR, and the effectiveness of PML with LWR on joint learning, which can be found in section \ref{sec:ablation}.

\subsection{Ablation Study of Convolutional Layers}\label{sec:cnn_layers}
In this section, we show the effect of convolutional layers and skip connection in Table \ref{table:layers}. 
\begin{table}[htbp]
\centering
\resizebox{0.8\columnwidth}{!}{
\begin{tabular}{c|ccc}
\hline
(\#shared, \#unshared, & \multicolumn{3}{c}{F1(\%) on MAPS+MDB-stem-synth}            \\
 skip-connection)               & Pitch & Onset & Onset\&Offset \\ \hline
OAF-retrain  &82.7 &83.8 &71.9 \\ \hline
(10, 4, True)       & 80.4   & 81.0   & 62.0  \\ \hline
(6, 4, True)       & 87.2   & 88.0   & 76.6  \\ \hline
(6, 4, False)                & 87.1  & 87.9  & 76.6  \\ \hline
(6, 8, True)       & \textbf{87.6}  & \textbf{88.3}  & \textbf{77.4} \\  \hline
(6, 8, False)              & 87.2  & 88.1  & 77.0  \\ \hline
(6, 12, True)     & 87.2   & 88.3   & 76.9  \\ \hline
(6, 12, False)             & 87.2  & 87.7  & 76.4  \\ \hline
\end{tabular}
}
\caption{Comparison of different numbers of convolutional layers and whether to use skip connection in terms of F1 score. These models is trained on mixed train dataset of MDB-stem-synth and MAPS. The test dataset is 1:1 mixed SP and MP data. skip-connection means whether to use skip connection. }
\label{table:layers}
\end{table}
We adjust the number of convolutional layers in shared layers and sub-task's stacks. All models use PML with LWR. Compared to third line model (6, 4, True), the second line model (10, 4, True) adds four convolutional layers in shared layers, while the performance at Pitch, Onset and Onset\&Offset metrics all decrease. This result indicates that oversharing decreases the discriminate ability of sub-tasks, and we need to limit the depth of shared layers. Compared to the model without skip connection, model with skip connection gets better performance with the same convolutional layers. This is because skip connection can utilize multi-level features and improve the performance. The fifth line model (6, 8, True) gets the best performance, and we adopt the configuration in other experiments.

\subsection{Ablation Study of Fusion Features}\label{sec:fusion_features}
We evaluate the effect of different fusion methods in this section, and the results are shown in Table \ref{table:fusion}. 
The model in last line, whose pitch detector does not use the output of Onset stack and Offset stack. 
From third line to sixth line, these model's onset prediction and offset prediction fuse the output of pitch stack as input features with the corresponding weights, while pitch prediction uses onset and offset features as JEPOO. 
The model in the second line is JEPOO.
Based on JEPOO, the model in the first line also fuses the onset prediction to offset, and the offset prediction to onset.
\begin{table}[htbp]
\centering
\resizebox{0.8\columnwidth}{!}{
\begin{tabular}{l|ccc}
\hline
\multirow{2}{*}{Fusion Methods}& \multicolumn{3}{c}{F1(\%) on MAPS+MDB-stem-synth}            \\
                      & Pitch & Onset & Onset\&Offset \\ \hline
Fuse features     & \multirow{2}{*}{87.1}  & \multirow{2}{*}{87.9}  & \multirow{2}{*}{75.0} \\  
for pitch, onset/offset \\ \hline
Fuse features     & \multirow{2}{*}{\textbf{87.6}}  & \multirow{2}{*}{\textbf{88.3}}  & \multirow{2}{*}{\textbf{77.4}} \\  
for pitch \\ \hline
Fuse features      & \multirow{2}{*}{87.0}  & \multirow{2}{*}{87.8}  & \multirow{2}{*}{75.2}  \\
for all sub-tasks(0.3) \\ \hline
Fuse features      & \multirow{2}{*}{86.4}  & \multirow{2}{*}{87.6}  & \multirow{2}{*}{73.7}  \\
for all sub-tasks(0.5) \\ \hline 
Fuse features      & \multirow{2}{*}{83.8}  & \multirow{2}{*}{84.8}  & \multirow{2}{*}{63.9}  \\
for all sub-tasks(0.8) \\ \hline
Fuse features      & \multirow{2}{*}{81.7}  & \multirow{2}{*}{83.9}  & \multirow{2}{*}{62.5}  \\ 
for all sub-tasks(1.0) \\ \hline
Without fusion    & 86.9   & 88.0   & 74.7  \\ \hline
\end{tabular}
}
\caption{Comparison of different fusion methods in terms of F1 score. These models is trained on mixed train dataset of MAPS and MDB-stem-synth. The test dataset is 1:1 mixed SP and MP data.}
\label{table:fusion}
\end{table}

From this table, we can conclude that the output of pitch stack may be the noise of onset and offset predictions. Because the performance of the fifth line model drops significantly than the last line model. Moreover, the performance increases as the weight of the pitch stack output decreases. Besides, our fusion method achieves the best performance. This may be because the majority of the pitch is not the onset/offset, and feeding such predictions to the predictor of onset/offset will actually harm the performance (making the model tend to always predict no onset/offset). On the other hand, the onset/offset frames indicate the beginning and the ending of pitches, so the onset/offset frames provide helpful information for the pitch prediction.

Ideally, the model should learn to ignore the noise of pitches in the onset and offset tasks if we fuse the feature of pitches, but this is only true when there is sufficiently labeled data. In reality, due to the nature of the data as described in Section~\ref{sec:intro}, there are much more frames with pitch labels than the frames with onset/offset labels, and this makes it difficult for the model to learn the patterns.

\end{document}